\documentclass[a4paper,12pt]{article}
\usepackage{amssymb} 
\usepackage{amsmath,amsfonts,epsfig,harvard}

\begin{document}
\title{On interface waves in misoriented pre-stressed 
incompressible elastic solids}

\author{Michel Destrade}
\date{2005}
\maketitle  

\bigskip

\begin{abstract}
\noindent
Some relationships, fundamental to the resolution of interface wave
problems, are presented. 
These equations allow for the derivation of explicit secular 
equations for problems involving waves localized near the plane 
boundary of anisotropic elastic half-spaces, 
such as Rayleigh, Sholte, or Stoneley waves. 
They are obtained rapidly, without recourse to the Stroh formalism.
As an application, the problems of Stoneley wave propagation and
of interface stability for misaligned predeformed incompressible
half-spaces are treated. 
The upper and lower half-spaces are made of the
same material, subject to the same prestress, 
and are rigidly bonded along a common principal plane. 
The principal axes in this plane do not however
coincide, and the wave propagation is studied in the direction of the
bisectrix of the angle between a principal axis of the upper half-space
and a principal axis of the lower half-space.

\end{abstract}

\newpage

\section{Introduction}

We all know from experience or from intuition that when we press 
together two large, well-polished, glass plates, they will stick 
together extremely well; 
in fact we will have a tough job trying to separate them again because 
in effect, the two glass panes have become one.
A somewhat similar process of ``gluing without glue'' is used in the 
microelectronics and optoelectronics industries to bond together 
\textit{semiconductor wafers} (G\"osele and Tong, 1998). 
When brought into contact, mirror-polished, flat, clean wafers made of 
almost any material are attracted \textit{via} Van der Walls forces 
and adhere in a rigid and permanent way.
This method of direct bonding allows for new and promising designs for 
insulators, sensors, actuators, nonlinear optics, light-emitting 
diodes, etc. 
Solid polymers can also be brought into permanent and rigid contact to 
manufacture \textit{polymer composites}. 
The main traditional technologies of polymer joining are:
mechanical fastening (bolts, rivets, fit joints) and:
adhesive bonding. 
Another technology, ``fusion bonding'', presents great advantages 
over the previous ones such as, avoidance of high stress 
concentrations, reduced surface treatment, less inhomogeneities at the 
interfaces, etc. 
A recent book by Ageorges and Ye (2002) presents a comprehensive 
description of fusion bonding, defined as ``the joining of two 
polymer parts by the fusion and consolidation of their interface''; 
in particular four classes of fusion bonding are listed: 
``bulk heating (co-consolidation, hot-melt adhesives, dual-resin 
bonding), frictional heating (spin welding, vibration welding, 
ultrasonic welding), electromagnetic heating (induction welding, 
microwave heating, dielectric heating, resistance welding), and 
two-stage techniques (hot plate welding, hot gas welding, radiant 
welding).''

For semiconductor wafer bonding, the most common and most economical 
combination is the silicon/silicon wafer bonding. 
Mozhaev \textit{et al.} (1998) considered the theoretical implications 
of misorientation when two identical silicon wafers are rigidly 
bonded. 
Specifically they considered, within the framework of anisotropic 
\textit{linear elasticity}, the propagation of interface (Stoneley) 
waves along the bisectrix of a twist angle of misorientation . 
This paper deals with the propagation of  waves of similar nature 
but within the framework of small motions superimposed on large static 
deformations (stress-induced anisotropy) in \textit{nonlinear 
elasticity}, which is required to describe the possible large elastic 
deformations of polymers (elastomers).
Specifically, two half-spaces made of the same hyperelastic 
incompressible material are maintained in the same static state 
of pure homogeneous deformation and are then rigidly bonded 
along a common principal plane of deformation, but in such a way that 
the two principal axes defining this plane for one half-space do not 
coincide with the two counterpart principal axes for the other 
half-space. 

The study of a superimposed infinitesimal interface wave propagating 
along the bisectrix of the misalignement angle provides insights into 
the possible ultrasonic non-destructive evaluation of the bond and of 
the angle of twist, and into the influence of the pre-strain on the 
interface stability.
Indeed in general, an increasingly tensile load applied on the 
semi-infinite bodies leads to faster speeds for an interfacial wave 
whereas a compressive load slows the wave down, until eventually 
the buckling/bifurcation criterion is met at the critical load, 
where the speed is zero.
Such investigations of interface waves and interfacial stability for 
two bonded hyperelastic pre-strained half-spaces are quite rare in the 
finite elasticity literature. 
They were initiated by Biot (1963), followed by a handful of papers:
Chadwick \& Jarvis (1979$a,b$), Dunwoody \& Villaggio (1988), 
Dowaikh \& Ogden (1991), Chadwick (1995), Destrade (to appear, $a$). 
Note however that in these articles, the two semi-infinite deformed 
bodies are always aligned so that all three principal axes of 
pre-deformation for the upper half-space  coincide with those for the 
lower half-space. 
Moreover, the interface wave propagates along one of the common 
principal axes, except for Chadwick \& Jarvis (1979$a,b$), who 
consider non-principal directions of propagation but for a specific 
simple form of strain energy function 
(compressible neo-Hookean materials). 

Some papers have been devoted to the study of non-principal interface 
waves (e.g. Flavin, 1963; Willson, 1973, 1974; 
Chadwick \& Jarvis, 1979$a,b,c$; Connor \& Ogden, 1995; 
Rogerson \& Sandiford, 1999; etc.) but explicit secular equations 
for tri-axially pre-strained materials have been found only in the 
case of neo-Hookean (compressible or incompressible) materials. 
This strain energy density is exceptional with respect to the 
propagation of inhomogeneous plane waves because for any direction 
of propagation in a principal plane 
(with associated orthogonal attenuation), the in-plane 
strain components always decouple from the anti-plane strain 
components.  
Section 2 presents the incremental equations of motion for 
non-principal interface waves in a tri-axially pre-strained material 
with a generic (non neo-Hookean) strain energy function, 
formulated as a first-order differential system for the 
six-component displacement-traction vector. 
It also presents the ``effective boundary conditions'', 
consequences of the continuity of the mechanical displacements and 
tractions across the interface.  
Namely, these conditions are that at the boundary, either one 
displacement and two tractions are zero, or two displacements and one 
traction are zero. 
Mozhaev \textit{et al.} (1998) noted this important result for bonded 
silicon/silicon wafers in linear anisotropic crystallography. 
It is further proved here that once the displacement-traction vector 
is normalized with respect to one of its non-zero components, then 
the normalized components are either real or pure imaginary 
(see Appendix for details). 
This Author recently obtained, in a quick and simple manner, some 
equations which are fundamental to the resolution of general interface 
boundary value problems (Destrade, 2003; Destrade, to appear, $b$) 
including Rayleigh (solid/vacuum interface), 
Scholte (solid/fluid interface), 
and Stoneley (solid/solid interface) waves;
they are presented in Subsection 3.1. 
The next Subsection brings together these fundamental equations and 
the effective boundary conditions for the title problem, with an 
explicit form of the secular equation that is, a polynomial of which 
the interface wave speed is a root.  
Finally Section 4 shows how numerical results can be obtained from the 
analysis, with the case of deformed semi-infinite bodies made of 
Mooney-Rivlin material, a model often used to describe the behaviour 
of incompressible rubber in large deformations.

\begin{figure}
\centering
\mbox{\epsfig{figure=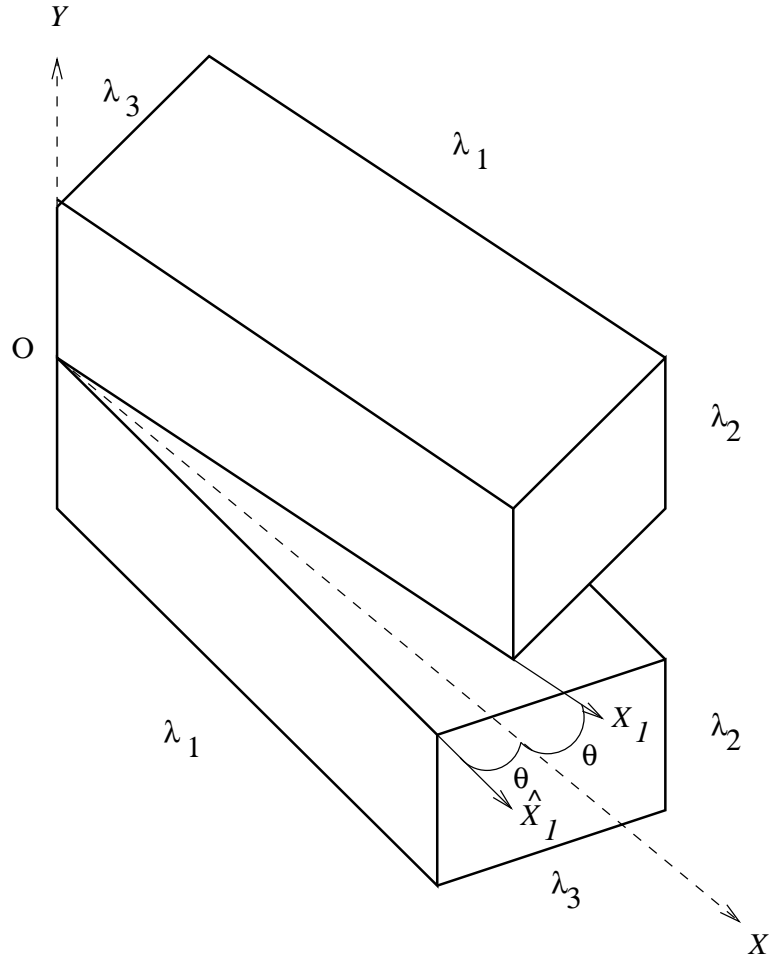}}
 \caption{Misoriented deformed cuboids at the interface}
\end{figure}

\section{Basic equations}
%
\subsection{Finite static pre-deformation}
%

Consider an infinite body made of two pre-strained hyperelastic 
semi-infinite bodies, rigidly bonded along a plane 
interface. 
Take the origin $O$ of a rectangular Cartesian coordinate system 
($OXYZ$) to lie in the boundary, so that the plane of separation 
between the half-spaces is $Y=0$. 
The upper half-space $Y \ge 0$ is made of an incompressible 
isotropic hyperelastic body, with mass density $\rho$ and strain 
energy function $W$, which has been subjected to a finite static pure 
homogeneous deformation with principal stretches 
$\lambda_1$, $\lambda_2$, $\lambda_3$  
($\lambda_1 \ne \lambda_2 \ne \lambda_3 \ne \lambda_1$ and 
$\lambda_1 \lambda_2 \lambda_3=1$) say, along the 
principal axes $OX_1X_2X_3$ say, such that  $X_2$ is aligned with 
$Y$ but $X_1$ ($X_3$) makes an angle $\theta$ with $X$ ($Z$). 
The lower half-space $Y \le 0$ is made of the \textit{same} body, 
which has been subjected to the \textit{same} pre-deformation that is, 
a pure homogeneous deformation   
with principal stretches $\lambda_1$, $\lambda_2$, $\lambda_3$ along 
the  principal axes $O\hat{X}_1\hat{X}_2\hat{X}_3$ (say) where 
$\hat{X}_2$ is aligned with $Y$, but such that now $\hat{X}_1$ 
($\hat{X}_3$) makes an angle $-\theta$ with $X$ ($Z$). 
Figure 1 summarizes this set-up with the representation of two 
parallelepipeds, one above $Y=0$, one below, which were unit cubes
 before the static deformation took place. 
%
\subsection{Incremental equations of motion and effective boundary 
conditions}
%

Now a small-amplitude inhomogeneous plane wave is superimposed 
upon the primary large static deformation. 
The wave propagates with speed $v$ and wave number $k$ in the $X$ 
direction and vanishes away from the interface $Y=0$. 
In other words, the corresponding incremental mechanical displacement 
$\mathbf{u}$ is of the form, 
\begin{equation} \label{wave} 
\mathbf{u} (x,y,z,t) 
  = \mathbf{U}(ky) e^{ik(x -vt)}, \quad 
\mathbf{U}(\pm \infty) = \mathbf{0},  
\end{equation} 
where $(Oxyz) = (OXYZ)$ is the rectangular Cartesian coordinate system 
associated with the motion. 

Similarly, the incremental nominal tractions $s_{j2}$ ($j=1,2,3$) 
acting on the planes $y = $ const. are of the form,  
\begin{equation} \label{tractions} 
s_{j2} (x,y,z,t) 
  = ik t_j(ky) e^{ik(x -vt)}, \quad 
t_j(\pm \infty) = 0.  
\end{equation} 

The following quantities allow for a compact form of the incremental 
equations of motion (Rogerson and Sandiford, 1999),
\begin{align} \label{gammaBeta}
& \gamma_{ij} := 
  (\lambda_i W_i-\lambda_j W_j)\lambda_i^2/(\lambda_i^2-\lambda_j^2)
  = \gamma_{ji} + \lambda_i W_i - \lambda_j W_j, 
\nonumber \\ 
& 2 \beta_{ij} := \lambda_i^2 W_{ii} 
   - 2 \lambda_i \lambda_j W_{ij} + \lambda_j^2 W_{jj}
 +  2(\lambda_i W_j-\lambda_j W_i)
         \lambda_i \lambda_j/(\lambda_i^2-\lambda_j^2) 
                = 2 \beta_{ji}, 
\end{align} 
(where $W_i := \partial W / \partial \lambda_i$) and also, 
\begin{align}
& c_\theta := \cos \theta, 
 &&  s_\theta := \sin \theta, 
\nonumber \\
& \eta_\theta :=  
    2 c_\theta^2(\beta_{12} + \gamma_{21}) + s_\theta^2 \gamma_{31}, 
&&  \nu_\theta := 
  c_\theta^2(\gamma_{12} - \gamma_{21}) 
   + s_\theta^2(\gamma_{32} - \gamma_{23}), 
\nonumber \\ 
& \mu_\theta :=  
    c_\theta^2 \gamma_{13} +  2 s_\theta^2 (\beta_{23} + \gamma_{23}), 
&&  \kappa_\theta :=   
       c_\theta s_\theta (\beta_{13} - \beta_{12} - \beta _{23}
                                        - \gamma_{21} - \gamma_{23}). 
\end{align}

Then a careful reading of the literature on incremental motions in 
incompressible materials (e.g. Ogden, 1984; Chadwick \textit{et al.}, 
1985; Chadwick, 1995; Rogerson \& Sandiford, 1999; Destrade \& Scott, 
to appear) and a fair amount of algebra reveal that the incremental 
equations of motion (in the upper half-space) can be cast in the form, 
\begin{equation}  \label{motion}
\mbox{\boldmath $\xi$}'= i \mathbf{N}\mbox{\boldmath $\xi$}, 
\quad \text{where} \quad 
\mbox{\boldmath $\xi$}(ky) = 
 [\mathbf{U}(ky), \mathbf{t}(ky)]^\text{T}, 
\end{equation}
the prime denotes differentiation with respect to $ky$, 
and $\mathbf{N}$ is the following matrix,  
\begin{equation}  \label{N}
  \mathbf{N}    = 
  \begin{bmatrix}
       0 & -c_\theta & 0 &  \frac{1}{\gamma_{21}} & 0 & 0 
       \\ 
      -c_\theta & 0 & -s_\theta & 0 & 0 & 0 
      \\  
       0 & -s_\theta & 0 &  0 & 0 &  \frac{1}{\gamma_{23}} 
      \\ 
       X - \eta_\theta & 0 & \kappa_\theta & 0 & -c_\theta & 0  
      \\  
       0 & X - \nu_\theta & 0 & -c_\theta & 0 & -s_\theta  
     \\  
       \kappa_\theta &     0     & X - \mu_\theta  
        & 0 & -s_\theta  & 0
  \end{bmatrix}, 
 \quad X := \rho v^2. 
\end{equation} 
The block structure of this matrix is reminescent of that for the 
``fundamental elasticity matrix'' $\mathbf{N}$ of linear anistropic 
elasticity (Ingebrigsten and Tonning, 1969), although its components 
are different.
In the lower half-space the equations of motion  are the same, 
allowing for the change in sign of $s_\theta$ and $\kappa_\theta$.

At the interface, the displacement-traction vector is continuous, 
\begin{equation} 
\mbox{\boldmath $\xi$}(0^+) = \mbox{\boldmath $\xi$}(0^-) 
 =: \mbox{\boldmath $\xi$}(0). 
\end{equation} 
Lengthy, but straightforward to establish, expressions for these 
quantities lead to the \textit{effective boundary conditions} at the 
interface. 
In effect, and as displayed by Mozhaev \textit{et al} (1998) in the 
linear anisotropic elasticity case, either one displacement component
and two tractions are zero at the interface, or \textit{vice-versa}. 
More specifically, and these details were not noted by 
Mozhaev \textit{et al} (1998),  $\mbox{\boldmath $\xi$}(0)$ must be of 
one of the two following forms. 
Either 
\begin{equation} \label{IAW1} 
\mbox{\boldmath $\xi$}(0) = 
U_2(0) [0,1,i\alpha_2, \beta_1,0,0]^\text{T}, 
\end{equation} 
where $\alpha_2$ and $\beta_1$ are real, 
or 
\begin{equation} \label{IAW2} 
\mbox{\boldmath $\xi$}(0) = 
U_1(0) [1,0,0,0, \alpha_1, i\beta_2]^\text{T}, 
\end{equation} 
where $\alpha_1$ and $\beta_2$ are real. 
The Appendix presents the derivation of these expressions. 
Following Mozhaev \textit{et al} (1998), the interface acoustic wave
satisfying \eqref{IAW1} (resp. \eqref{IAW2}) is called IAW1 
(resp. IAW2).
   
%
\section{Explicit resolution}
%
\subsection{Fundamental equations for interface boundary value 
problems}

Here a quick derivation is made of some equations 
(Destrade, 2003) which are fundamental to the resolution of 
interface boundary problems involving waves or static deformations 
which are localized near, or equivalently vanish away from, the plane 
interface of a semi-infinite body. 
The equations are valid for unconstrained or constrained, 
pre-deformed nonlinearly elastic or anisotropic linearly 
elastic, materials. 
They do not rely on the Stroh (1958) formalism. 
All that is required for their derivation are the few manipulations, 
available in the literature, leading to the equations of motion and 
boundary conditions in the form \eqref{motion}, \eqref{wave}$_2$, 
\eqref{tractions}$_2$, that is 
\begin{equation} \label{motion-BC}
\mbox{\boldmath $\xi$}'= i \mathbf{N}\mbox{\boldmath $\xi$}, 
\quad  
\mbox{\boldmath $\xi$}(\infty) = \mathbf{0}. 
\end{equation}

In general here, $\mathbf{N}$ is a square matrix of even dimensions, 
$2p \times 2p$ say. 
Now a simple induction process (Currie, 1979; Ting, 2003) shows 
that $\mathbf{N}^n$, where $n$ is any positive or negative integer, 
is of the following form (see \eqref{N} for $n=1$), 
\begin{equation} 
\mathbf{N}^n = 
\begin{bmatrix} 
 \mathbf{N_1}^\text{(n)} &  \mathbf{N_2}^\text{(n)}  \\ 
 \mathbf{K}^\text{(n)}     &  \mathbf{N_1}^\text{(n)T}
\end{bmatrix}, 
\quad \text{with}  \quad 
 \mathbf{K}^\text{(n)} =  \mathbf{K}^\text{(n)T}, \quad
 \mathbf{N_2}^\text{(n)}  =  \mathbf{N_2}^\text{(n)T}. 
\end{equation}  
Consequently, pre-multiplication of $\mathbf{N}^n$ by 
$\mathbf{\hat{I}}$, defined as 
\begin{equation} 
\mathbf{\hat{I}} := 
\begin{bmatrix} 
 \mathbf{0}    &  \mathbf{I_p}  \\ 
 \mathbf{I_p} &  \mathbf{0}
\end{bmatrix}, 
\quad 
\text{where }  \mathbf{I_p}    \text{ is the }
 p \times p \text{ identity matrix}, 
\end{equation}  
leads to a symmetric matrix, 
\begin{equation} 
\mathbf{\hat{I}} \mathbf{N}^n = 
\begin{bmatrix} 
 \mathbf{K}^\text{(n)} &  \mathbf{N_1}^\text{(n)T}  \\ 
 \mathbf{N_1}^\text{(n)}     &  \mathbf{N_2}^\text{(n)}
\end{bmatrix} 
= (\mathbf{\hat{I}} \mathbf{N}^n)^\text{T}. 
\end{equation}  

Now take the scalar product of \eqref{motion-BC}$_1$ by 
$ \mathbf{\hat{I}} \mathbf{N}^n 
     \overline{\mbox{\boldmath $\xi$}}$ and add the complex conjugate 
quantity to obtain 
$\mbox{\boldmath $\xi$}' \mathbf{ \cdot \hat{I}} \mathbf{N}^n 
     \overline{\mbox{\boldmath $\xi$}} + 
 \mbox{\boldmath $\xi$} \mathbf{ \cdot \hat{I}} \mathbf{N}^n 
     \overline{\mbox{\boldmath $\xi$}}' =0$. 
Direct integration between zero and infinity yields, using 
\eqref{motion-BC}$_2$, the \textit{fundamental equations} sought, 
\begin{equation} \label{fundamental} 
\mbox{\boldmath $\xi$}(0) \mathbf{ \cdot \hat{I}} \mathbf{N}^n 
     \overline{\mbox{\boldmath $\xi$}}(0) = 0. 
\end{equation}  
The Cayley-Hamilton theorem states that $\mathbf{N}$ satisfies its 
own characteristic polynomial of degree $2p$. 
Hence there are only $2p-1$ powers of $\mathbf{N}$ which are linearly 
independent and so, Eqs. \eqref{fundamental} generate at most $2p-1$ 
linearly independent equations. 
For instance in our case $\mathbf{N}$ is a $6 \times 6$ matrix, 
leading to only 5 linearly independent equations. 
These are nevertheless enough to deduce the secular equation 
explicitly, as is now seen. 
 
\subsection{Secular equation}

The fundamental equations \eqref{fundamental} now prove useful in the 
resolution of the paper's problem. 
The structure of $\mathbf{\hat{I}} \mathbf{N}^n$ depends on the parity 
of $n$. 
For instance, 
\begin{equation} 
\mathbf{\hat{I}} \mathbf{N}^n = 
\begin{bmatrix} 
     0       &N^n_{42}&      0      &N^n_{11}&      0      &N^n_{31}
 \\ 
N^n_{42}&       0     &N^n_{53}&      0      &N^n_{22}&    0        
 \\ 
     0       &N^n_{53}&      0      &N^n_{13}&      0      &N^n_{33}
 \\ 
N^n_{11}&       0     &N^n_{13}&      0      &N^n_{15}&    0        
 \\ 
     0       &N^n_{22}&      0      &N^n_{15}&      0      &N^n_{26}
 \\ 
N^n_{31}&       0     &N^n_{33}&      0      &N^n_{26}&    0         
 \end{bmatrix}, 
 \quad \text{for } n = -2, 2, 
\end{equation}  
where $N^n_{ij} := (\mathbf{N}^n)_{ij}$; thus for example, 
$ N^2_{42} := (\mathbf{N}^2)_{42} = N_{4k} N_{k2}$. 
The forms \eqref{IAW1} and \eqref{IAW2} for 
$\mbox{\boldmath $\xi$}(0)$ lead to two trivial identities when 
\eqref{fundamental} are written at $n = -2,2$.
On the other hand,  
\begin{equation} 
\mathbf{\hat{I}} \mathbf{N}^n = 
\begin{bmatrix} 
 N^n_{41}&       0     &N^n_{43}&      0      &N^n_{21}&    0        
 \\ 
     0       &N^n_{52}&      0      &N^n_{12}&      0      &N^n_{32}
 \\ 
N^n_{43}&       0     &N^n_{63}&      0      &N^n_{23}&    0        
 \\ 
     0       &N^n_{12}&      0      &N^n_{14}&      0      &N^n_{16}
 \\ 
N^n_{21}&       0     &N^n_{23}&      0      &N^n_{25}&    0        
 \\ 
    0       &N^n_{32}&      0      &N^n_{16}&      0      &N^n_{36} 
 \end{bmatrix}, 
 \quad \text{for } n = -1, 1, 3,   
\end{equation}  
and so, the fundamental equations \eqref{fundamental}, written for 
IAW1 that is, $\mbox{\boldmath $\xi$}(0)$ given by \eqref{IAW1}, 
and for $n = -1,1,3$,  yield the following system, 
\begin{equation} \label{system1}
  \begin{bmatrix} 
 N^{-1}_{12} & N^{-1}_{14} & N^{-1}_{63}       \\ 
 N^1_{12}    & N^1_{14}     & N^1_{63}       \\ 
 N^3_{12}    & N^3_{14}     & N^3_{63}       
  \end{bmatrix} 
  \begin{bmatrix} 
   2\beta_1         \\ 
     \beta_1^2      \\ 
    \alpha_2^2       
 \end{bmatrix} 
  =  \begin{bmatrix} 
       -N^{-1}_{52}  \\ 
       -N^1_{52}  \\ 
       -N^3_{52}       
  \end{bmatrix}. 
\end{equation}  
When the fundamental equations \eqref{fundamental} are written for 
IAW2 that is, $\mbox{\boldmath $\xi$}(0)$ given by \eqref{IAW2}, 
and for $n = -1,1,3$, they yield the system, 
\begin{equation} \label{system2}
  \begin{bmatrix} 
 N^{-1}_{21} & N^{-1}_{25} & N^{-1}_{36}       \\ 
 N^1_{21}    & N^1_{25}     & N^1_{36}       \\ 
 N^3_{21}    & N^3_{25}     & N^3_{36}       
  \end{bmatrix} 
  \begin{bmatrix} 
   2\alpha_1         \\ 
     \alpha_1^2      \\ 
    \beta_2^2       
 \end{bmatrix} 
  =  \begin{bmatrix} 
       -N^{-1}_{41}  \\ 
       -N^1_{41}  \\ 
       -N^3_{41}       
  \end{bmatrix}. 
\end{equation}  
These equations are of the type $M_{ik}b_k = a_i$ 
where here $\mathbf{M}$ is the $3 \times 3$ matrix on the left 
hand-side of \eqref{system1} for IAW1 and of \eqref{system2} for IAW2, 
$\mathbf{a}$ is the vector on the right hand-side, and 
$\mathbf{b} = [2\beta_1, \beta_1^2,  \alpha_2^2]^\text{T}$, 
$[2\alpha_1, \alpha_1^2,  \beta_2^2]^\text{T}$, for IAW1 and for IAW2, 
respectively. 
By Cramer's rule, the unique solution is $b_k = \Delta_k / \Delta$, 
where $\Delta  = \text{ det } \mathbf{M}$, and $\Delta_k$ is the 
determinant of the matrix obtained from $\mathbf{M}$ by replacing the 
$k$-th vector column with $\mathbf{a}$. 
However two components of $\mathbf{b}$ are related one to another. 
Specifically, $b_1^2 = 4b_2$, which is the \textit{explicit secular 
equation for interface waves propagating along the bissectrix of the 
misorientation angle  for two identical, rigidly bonded, pre-deformed, 
hyperelastic half-spaces}, 
\begin{equation} \label{secular}
\Delta_1^2 - 4\Delta \Delta_2 = 0. 
\end{equation}

%
\section{Numerical results for rubberlike materials}

The classical Mooney-Rivlin model has been used extensively to model 
the behaviour of incompressible rubberlike materials in large 
deformations.
Its strain energy function $W$ is given by 
\begin{equation} \label{MR}
 2W = C(\lambda_1^2 + \lambda_2^2 + \lambda_3^2 - 3) 
  + D(\lambda_1^2\lambda_2^2 + \lambda_2^2\lambda_3^2 
       + \lambda_3^2\lambda_1^2 - 3), 
\end{equation}
where $C$ and $D$ are constant material parameters. 
In that case, the $\gamma_{ij}$ and $\beta_{ij}$ defined in 
\eqref{gammaBeta} reduce to ($i \ne j \ne k \ne i$),
\begin{equation}
\gamma_{ij} = \lambda_i^2 (C + D\lambda_k^2), \quad 
2\beta_{ij} = (\lambda_i^2 + \lambda_j^2) (C + D\lambda_k^2)
            = \gamma_{ij} + \gamma_{ji}.
\end{equation}

Now the peculiarities of interface acoustic waves in deformed 
hyperelastic materials are highlighted. 
In particular, and in contrast with the corresponding situation in 
linear anisotropic elasticity, it is seen that there exist certain 
ranges of misorientation and certain ranges of pre-stretch ratios 
for the existence of an IAW. 
Also,  interface instability may arise at certain compressive critical 
ratios.

%
\subsection{Mooney-Rivlin material in tri-axial strain}
 
First, in order to make the connection with results obtained by 
Rogerson \& Sandiford (1999) about non-principal surface waves, 
the material constants $C$, 
$D$, and the stretch ratios $\lambda_i^2$ are fixed at the following 
values,
\begin{equation} 
C = 2.0, \quad D = 0.8, \quad 
\lambda_1^2 = 3.695,  \quad \lambda_2^2 = 0.7, 
  \quad \lambda_3^2 = 0.387. 
\end{equation}  

The secular equation \eqref{secular} is a polynomial of 
degree 10 in $X = \rho v^2$ for IAW1, of degree 6 for IAW2.
Out of the 16 possible roots, only one leads to an interface wave of 
the Stoneley type satisfying the known conditions that its speed is 
bounded above by the speed of the slowest homogeneous bulk wave and 
below by the speed of the Rayleigh surface wave associated with either 
half-space (Barnett \textit{et al.}, 1985). 
Moreover this wave, of the IAW1 type, satisfies these requirements 
only within a limited range of misorientation angle, approximatively 
$16.7^\text{o} < \theta < 73.6^\text{o}$.
This situation is in sharp contrast with the case of silicon/silicon 
wafers in linear anisotropic elasticity (Mozhaev \textit{et al.}, 1998)
where the IAW1 was found to exist for all $\theta$.

Figure 2 depicts the variations of the relevant root to the secular 
equation, scaled as $\sqrt{X} = \sqrt{\rho v^2}$, with $\theta$ 
(thick curve). 
The speeds of  two homogeneous shear waves 
(roots of $\text{det }\mathbf{N} = 0$) are represented by the two thin 
curves above, and the speed of the Rayleigh wave, by the thin curve 
below.

\begin{figure}
\centering
\mbox{\epsfig{figure=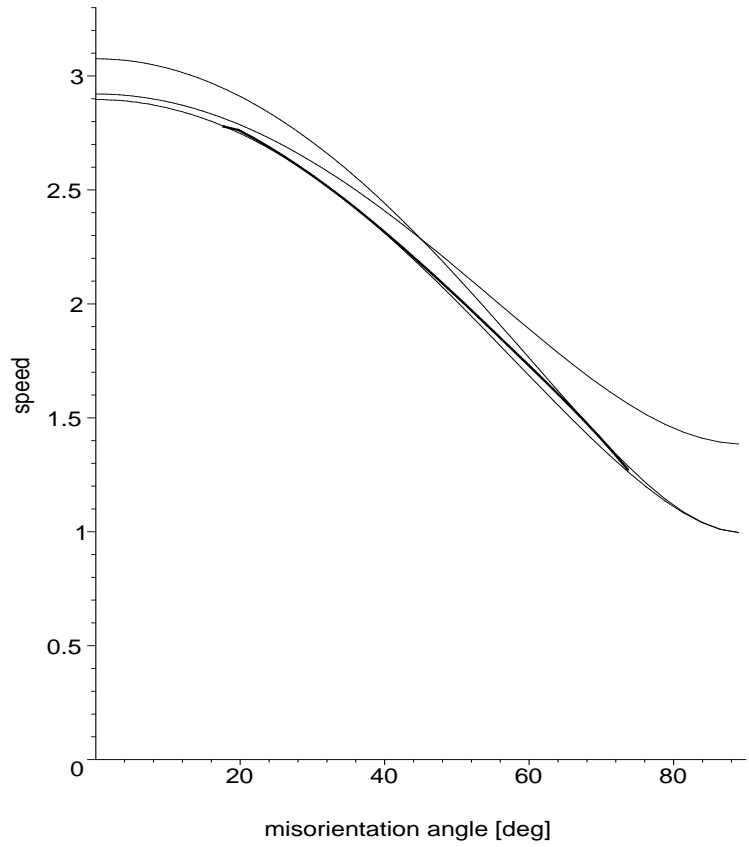, width=0.7\textwidth, 
height=0.8\textwidth}}
 \caption{Mooney-Rivlin bimaterial subject to the triaxial pre-stretch 
$\lambda_1^2 = 3.695,   \lambda_2^2 = 0.7, \lambda_3^2 = 0.387$: 
plot of the Stoneley wave speed with $\theta$ (thick curve), 
bonded above by the shear wave speeds 
(crossing at $\theta = 45^\text{o}$) and below by the Rayleigh wave 
speed.}
\end{figure}

%
\subsection{Mooney-Rivlin material in compressive plane strain}

Next, the half-spaces are assumed to have been pre-deformed in such 
a way that they were not allowed to expand in the $Z$-direction 
($\lambda_3 = 1$). 
For this example, the different parameters take the following 
numerical and algebraic values, 
\begin{equation} 
C = 2.0, \quad D = 0.8, \quad \theta = 30^\text{o}, \quad 
\lambda_1 = \lambda,  \quad \lambda_2 = \lambda^{-1}, 
  \quad \lambda_3 = 1. 
\end{equation}  

Figure 3 shows the variations of the speeds as functions of 
$\lambda$ in compression ($\lambda < 1$). 
The two upper dashed curves represent the speeds of homogeneous 
shear waves;
the intermediate thick curve, the speed of the Stoneley wave 
(type: IAW1); the bottom thin curve, the speed of the Rayleigh wave 
associated with either deformed half-space. 
Again, the situation is different from that encountered in linear 
anisotropic elasticity with silicon/silicon wafers 
(Mozhaev \textit{et al.}, 1998). 
In particular, the Stoneley wave exists only for stretch ratios  
greater than 0.321 and the Rayleigh wave for stretch ratios  
greater than 0.451. 
In between these two critical stretches, there is a range were 
the Stoneley wave exists but not the Rayleigh wave. 
At the critical stretch, instability might occur; 
hence is presented an example where two bonded deformed half-spaces 
exhibit interfacial stability at some  compressive stretch ratios 
for which the separated half-spaces are unstable (at least in the 
linearized theory).

\begin{figure}
\centering
\mbox{\epsfig{figure=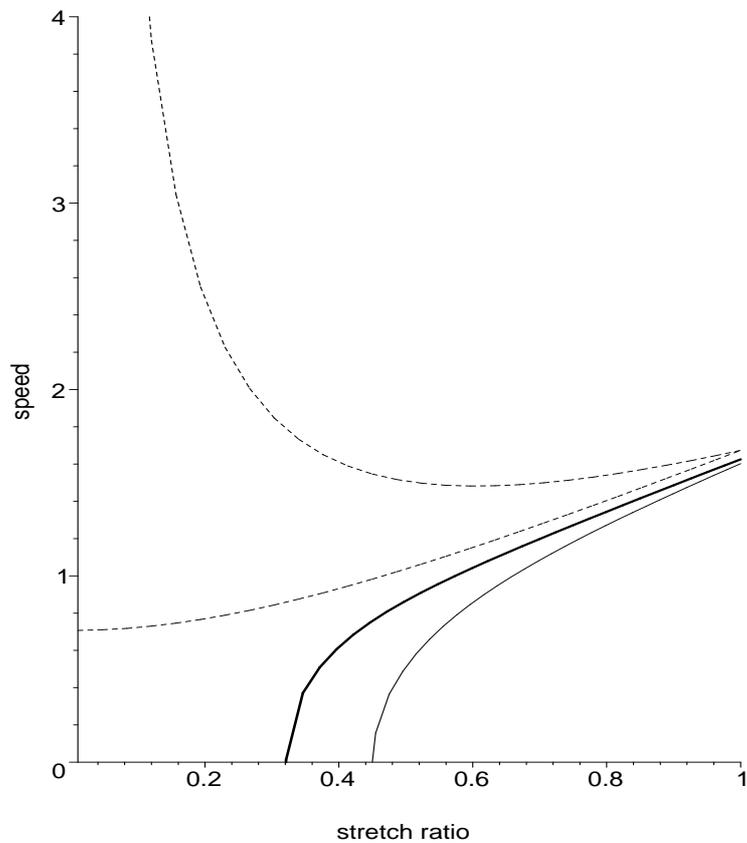, width=0.7\textwidth, 
height=0.8\textwidth}}
 \caption{Deformed Mooney-Rivlin bimaterial in compressive 
plane strain at $\theta = 30^\text{o}$: 
variations with $\lambda$ of the shear (dashed curves), 
Stoneley (thick curve), and Rayleigh (lower curve) wave speeds.}
\end{figure}


\appendix 
 \renewcommand{\thesection}{\Alph{section}}

\section*{Appendix: effective boundary conditions} 
\setcounter{section}{1} 
\setcounter{equation}{0} 
      
Here two main results about the boundary conditions of the paper's 
problem are established. 
The first result is that at the interface, either one displacement 
component and two tractions are zero, or \textit{vice-versa}. 
Mozhaev \textit{et al.} (1998) found this result for the corresponding 
problem in anisotropic linear elasticity; because of space limitations 
they only sketched the proof.
The second result is that once the displacement-traction vector at the 
interface is normalized with respect to one of the three non-zero 
components, then the two others are either purely real or purely 
imaginary quantities.
This Appendix is a generalization of results obtained by Ting 
(to appear) for surface (Rayleigh) waves and by Destrade (2003) for 
2-partial interface (Stoneley) waves.

Consider, for the \textit{upper half-space}, a combination of 
inhomogeneous plane waves of the form, 
\begin{equation}
\mbox{\boldmath $\xi$} = 
 \mbox{\boldmath $\xi^i$} e^{ik(x + p_i y - vt)}, 
\end{equation} 
where $\mbox{\boldmath $\xi^i$}$ is a constant vector and 
$\Im(p_i)>0$ to ensure decay of the wave with distance from the 
interface. 
Substitution of this form of solution into the equations of motion 
\eqref{motion} shows that the $p_i$ are roots of the bicubic: 
$\text{det } (\mathbf{N} - p\mathbf{1})=0$, or 
(Rogerson and Sandiford, 1999),
\begin{equation} \label{bicubic} 
\gamma_{21} \gamma_{23} p^6
 - [(\gamma_{21} + \gamma_{23})X - c_1]p^4 
  + (X^2 - c_2X + c_3)p^2 
    + (X-c_4)(X-c_5) = 0, 
\end{equation} 
with
\begin{align} 
& c_1 =  (\gamma_{21}\gamma_{13} + 2\beta_{12}\gamma_{23})c_\theta^2 
        + (\gamma_{23}\gamma_{31} + 2\beta_{23}\gamma_{21})s_\theta^2, 
\nonumber \\ 
& c_2 =  (\gamma_{23} + \gamma_{13} + 2\beta_{12})c_\theta^2 
              + (\gamma_{21} + \gamma_{31} + 2\beta_{23})s_\theta^2,  
\nonumber \\ 
& c_3 =  (\gamma_{12}\gamma_{23} + 2\beta_{12}\gamma_{13})c_\theta^4 
        + (\gamma_{21}\gamma_{32} + 2\beta_{23}\gamma_{31})s_\theta^4  
\nonumber \\ 
& \phantom{123456}  + [\gamma_{12}\gamma_{21} + \gamma_{13}\gamma_{31}
                 +\gamma_{23}\gamma_{32}     
                   - (\beta_{13} - \beta_{12} - \beta_{23})^2
                    + 4 \beta_{12}\beta_{23}]c_\theta^2 s_\theta^2,  
\nonumber \\ 
& c_4 = \gamma_{12}c_\theta^2 + \gamma_{32}s_\theta^2,  
\nonumber \\ 
& c_5 =  \gamma_{13}c_\theta^4 + 2 \beta_{13}c_\theta^2s_\theta^2 
                   + \gamma_{31}s_\theta^4. 
\end{align} 

Let $p_1, p_2, p_3$  be the three roots with positive imaginary part. 
Then the wave combination is 
\begin{equation}
\mbox{\boldmath $\xi$} = 
 \gamma_1 \mbox{\boldmath $\xi^1$} e^{ik(x + p_1 y - vt)} + 
  \gamma_2 \mbox{\boldmath $\xi^2$} e^{ik(x + p_2 y - vt)} + 
   \gamma_3 \mbox{\boldmath $\xi^3$} e^{ik(x + p_3 y - vt)}, 
\quad (y \ge 0), 
\end{equation} 
where the $\gamma_i$ are constant numbers. 

In the \textit{lower half-space}, the equations of motion are of the 
same form as in the upper half-space with $\theta$ changed to 
$-\theta$. 
Hence only $s_\theta$ and $\kappa$ change signs in the components of 
$\mathbf{N}$ whilst the coefficients of the associated characteristic 
polynomial remain unchanged. 
Consequently, a suitable combination of inhomogeneous plane waves is 
here 
\begin{equation}
\mbox{\boldmath $\hat{\xi}$} = 
 \hat{\gamma}_1 \mbox{\boldmath $\hat{\xi}^1$} e^{ik(x - p_1 y - vt)}
 + 
  \hat{\gamma}_2 \mbox{\boldmath $\hat{\xi}^2$} e^{ik(x - p_2 y - vt)}
 + 
 \hat{ \gamma}_3 \mbox{\boldmath $\hat{\xi}^3$} e^{ik(x - p_3 y - vt)},
 \quad (y \le 0), 
\end{equation} 
where the $\hat{\gamma}_i$ are constant numbers and the 
$\mbox{\boldmath $\hat{\xi}^i$}$ constant vectors.

Now two possibilities arise for the roots $p_1, p_2, p_3$ of 
\eqref{bicubic}. 
Either 
\begin{equation} 
\text{(I) } p_i = i q_i \; (q_i>0), \quad \text{or} \quad 
\text{(II) } p_1 = - \overline{p_2}, \quad p_3 =i q_3 \; (q_3>0). 
\end{equation} 

In Case (I) it is a straightforward matter to show, using the adjoint 
of $(\mathbf{N} - p\mathbf{1})$, that the $\mbox{\boldmath $\xi^i$}$ 
(upper half-space) can be written in the form, 
\begin{equation} 
\mbox{\boldmath $\xi^i$} = [a_i, ib_i, c_i, id_i, e_i, if_i]^\text{T}, 
\end{equation} 
say, where $a_i, b_i, c_i, d_i, e_i, f_i$ ($i = 1,2,3$) are 
\textit{real} numbers. 
Furthermore, the vectors $\mbox{\boldmath $\hat{\xi}^i$}$ 
(lower half-space, where $\theta$ is changed to $-\theta$) 
are then in the form, 
\begin{equation} 
\mbox{\boldmath $\hat{\xi}^i$} 
    = [-a_i, ib_i, c_i, id_i, -e_i, -if_i]^\text{T}.
\end{equation} 
At the interface $y=0$ the displacement-traction vector is continuous, 
$\mbox{\boldmath $\xi$}(x,0,t) = \mbox{\boldmath $\hat{\xi}$}(x,0,t)$, 
or 
\begin{equation} 
\gamma_1 \mbox{\boldmath $\xi^1$}  + 
  \gamma_2 \mbox{\boldmath $\xi^2$} + 
   \gamma_3 \mbox{\boldmath $\xi^3$} =
\hat{\gamma}_1 \mbox{\boldmath $\hat{\xi}^1$} + 
 \hat{\gamma}_2 \mbox{\boldmath $\hat{\xi}^2$}  + 
  \hat{ \gamma}_3 \mbox{\boldmath $\hat{\xi}^3$}. 
\end{equation} 
These six equations are recast as 
\begin{equation}  \label{sys}
 \mathbf{A}
  (\mbox{\boldmath $\gamma$} + \mbox{\boldmath $\hat{\gamma}$})
  =  \mathbf{0}, \quad 
 \mathbf{B}
  (\mbox{\boldmath $\gamma$} - \mbox{\boldmath $\hat{\gamma}$})
  =  \mathbf{0}, 
\end{equation} 
where
\begin{equation} \label{AB}
  \mathbf{A} =  \begin{bmatrix}
                         a_1 & a_2 &  a_3  \\
                         e_1 & e_2 &  e_3  \\
                         if_1 & if_2 & if_3
                       \end{bmatrix}, \quad 
\mathbf{B} = \begin{bmatrix}
                         ib_1 & ib_2 &  ib_3  \\
                          c_1 &  c_2 &   c_3  \\
                         id_1 & id_2 & id_3
                       \end{bmatrix}, \quad 
\mbox{\boldmath $\gamma$} 
                   = \begin{bmatrix}
                           \gamma_1 \\
                           \gamma_2 \\ 
                           \gamma_3
                     \end{bmatrix}, \quad 
\mbox{\boldmath $\hat{\gamma}$} 
                   = \begin{bmatrix}
                           \hat{\gamma}_1 \\
                           \hat{\gamma}_2 \\ 
                           \hat{\gamma}_3
                     \end{bmatrix}.
\end{equation} 
For non-trivial solutions to exist,
 either (a)  $\text{det } \mathbf{A} = 0$ 
 or (b) $\text{det } \mathbf{B} = 0$. 
In case (a), Eq.~\eqref{sys}$_2$ leads to  
$ \mbox{\boldmath $\gamma$}
  = \mbox{\boldmath $\hat{\gamma}$}$. 
Then  Eq.~\eqref{sys}$_1$ reads 
\begin{equation} 
 \mathbf{A}  \mbox{\boldmath $\gamma$}= \mathbf{0}.
\end{equation}
Owing to the form \eqref{AB}$_1$ of $\mathbf{A}$, this condition is 
satisfied when the components of $\mbox{\boldmath $\gamma$}$ 
are all real. 
In conclusion, $\mbox{\boldmath $\xi$}(0)$ now reads 
\begin{equation} 
 \mbox{\boldmath $\xi$}(0) = 
\gamma_1 \begin{bmatrix}
                        0 \\ ib_1 \\ c_1 \\ id_1 \\ 0 \\ 0 
                  \end{bmatrix}
 + \gamma_2 \begin{bmatrix}
                         0 \\ ib_2 \\ c_2 \\ id_2 \\ 0 \\ 0 
                      \end{bmatrix}
  + \gamma_3 \begin{bmatrix}
                         0 \\ ib_3 \\ c_3 \\ id_3 \\ 0 \\ 0 
                      \end{bmatrix},  \quad
\gamma_i \text{ real}, 
\end{equation} 
that is, $\mbox{\boldmath $\xi$}(0)$ is of the form \eqref{IAW1}. 
A similar procedure shows that in case (b), 
$\mbox{\boldmath $\xi$}(0)$ is of the form \eqref{IAW2}.

In Case (II) the $\mbox{\boldmath $\xi^i$}$ 
(upper half-space) can be written in the form, 
\begin{equation} 
\mbox{\boldmath $\xi^1$} =  \begin{bmatrix}
                        a_1 \\ b_1 \\ c_1 \\ d_1 \\ e_1 \\ f_1 
                  \end{bmatrix}, \quad
\mbox{\boldmath $\xi^2$} =  \begin{bmatrix}
              \overline{a_1} \\  -\overline{b_1} \\  \overline{c_1} \\ 
             - \overline{d_1} \\  \overline{e_1} \\  -\overline{f_1} 
                  \end{bmatrix}, \quad
\mbox{\boldmath $\xi^3$} =  \begin{bmatrix}
                        a_3 \\ ib_3 \\ c_3 \\ id_3 \\ e_3 \\ if_3 
                  \end{bmatrix}, 
\end{equation} 
say, where $a_1, b_1, c_1, d_1, e_1, f_1$ are \textit{complex} 
and $a_3, b_3, c_3, d_3, e_3, f_3$ are \textit{real}. 
The vectors $\mbox{\boldmath $\hat{\xi}^i$}$ 
(lower half-space) are then in the form, 
\begin{equation} 
\mbox{\boldmath $\hat{\xi}^1$} =  \begin{bmatrix}
                        -a_1 \\ b_1 \\ c_1 \\ d_1 \\ -e_1 \\ -f_1 
                  \end{bmatrix}, \quad
\mbox{\boldmath $\hat{\xi}^2$} =  \begin{bmatrix}
              -\overline{a_1} \\  -\overline{b_1} \\ \overline{c_1} \\ 
             - \overline{d_1} \\  -\overline{e_1} \\  \overline{f_1} 
                  \end{bmatrix}, \quad
\mbox{\boldmath $\hat{\xi}^3$} =  \begin{bmatrix}
                        -a_3 \\ ib_3 \\ c_3 \\ id_3 \\ -e_3 \\ -if_3 
                  \end{bmatrix}.
\end{equation} 
The continuity of the displacement-traction vector at $y=0$ can again 
be written in the form \eqref{AB} where now 
\begin{equation} \label{AB(II)}
  \mathbf{A} =  \begin{bmatrix}
                         a_1 & \overline{a_1}  &  a_3  \\
                         e_1 & \overline{e_1}  &  e_3  \\
                          f_1 & - \overline{f_1} & if_3
                       \end{bmatrix}, \quad 
\mathbf{B} = \begin{bmatrix}
                         b_1 & - \overline{b_1} &  ib_3  \\
                         c_1 &   \overline{c_1} &   c_3  \\
                         d_1 & - \overline{d_1} &  d_3
                       \end{bmatrix}.
\end{equation} 
Again case (a)  $\text{det } \mathbf{A} = 0$ or case (b) 
$\text{det } \mathbf{B} = 0$ arise. 
In case (a),  
$ \mbox{\boldmath $\gamma$}
  = \mbox{\boldmath $\hat{\gamma}$}$  
and so 
 $\mathbf{A}  \mbox{\boldmath $\gamma$}= \mathbf{0}$.
Owing to the form \eqref{AB(II)}$_1$ of $\mathbf{A}$, this condition 
is satisfied when $\mbox{\boldmath $\gamma$}$ is of the form 
$ \mbox{\boldmath $\gamma$}
 = [\gamma_1,\overline{\gamma_1},\gamma_3]^\text{T}$, where 
$\gamma_1$ is complex and $\gamma_3$ real.
In conclusion, $\mbox{\boldmath $\xi$}(0)$ now reads 
\begin{equation} 
 \mbox{\boldmath $\xi$}(0) = 
\gamma_1 \begin{bmatrix}
                        0 \\ b_1 \\ c_1 \\ d_1 \\ 0 \\ 0 
                  \end{bmatrix}
 + \overline{\gamma_1} \begin{bmatrix}
 0 \\ -\overline{b_1} \\ \overline{c_1} \\ -\overline{d_1} \\ 0 \\ 0 
                      \end{bmatrix}
  + \gamma_3 \begin{bmatrix}
                         0 \\ ib_3 \\ c_3 \\ id_3 \\ 0 \\ 0 
                      \end{bmatrix},  \quad
\gamma_3 \text{ real}, 
\end{equation} 
which means that, once normalized, $\mbox{\boldmath $\xi$}(0)$ 
is of the form \eqref{IAW1}. 
A similar procedure shows that in case (b), 
$\mbox{\boldmath $\xi$}(0)$ is of the form \eqref{IAW2}.

\end{document}